\documentclass{article}

\usepackage{amssymb,amsfonts,amsmath}
\usepackage{cite,enumerate,float}
\usepackage{color}
\usepackage{tikz}
\usetikzlibrary{arrows,snakes,backgrounds}
\usepackage{url}
\usepackage{mathtools}
\usepackage[cmtip,all]{xy}

\def\be{\begin{eqnarray}}
\def\ee{\end{eqnarray}}
\def\nn{\nonumber}

\def\p{\partial}

\def\wc{weak composition\ }
\def\wcs{weak compositions\ }

\definecolor{red}{rgb}{1,0,0}
\definecolor{orange}{rgb}{1,0.5,0}
\definecolor{violet}{rgb}{0.7,0,1}

\newcommand{\ttop}[1]{
  q^{\hat{D}_#1}
}

\newcommand{\longsquiggly}{\xymatrix{{}\ar@{~>}[r]&{}}}

\hoffset= - 1.0in         % switch off for draft style
\begin{document}

\title{\vspace{1.5cm}\bf
Symmetric polynomials:
DIM integrable systems\\
versus twisted Cherednik systems
}

\author{
A. Mironov$^{b,c,d,}$\footnote{mironov@lpi.ru,mironov@itep.ru},
A. Morozov$^{a,c,d,}$\footnote{morozov@itep.ru},
A. Popolitov$^{a,c,d,}$\footnote{popolit@gmail.com}
}

\date{ }

\maketitle

\vspace{-6cm}

\begin{center}
  \hfill MIPT/TH-04/26\\
  \hfill FIAN/TD-04/26\\
  \hfill ITEP/TH-04/26\\
  \hfill IITP/TH-04/26
\end{center}

\vspace{4.5cm}

\begin{center}
$^a$ {\small {\it MIPT, Dolgoprudny, 141701, Russia}}\\
$^b$ {\small {\it Lebedev Physics Institute, Moscow 119991, Russia}}\\
$^c$ {\small {\it NRC ``Kurchatov Institute", 123182, Moscow, Russia}}\\
$^d$ {\small {\it Institute for Information Transmission Problems, Moscow 127994, Russia}}
\end{center}

\vspace{.1cm}

\begin{abstract}
We discuss interrelations between eigenfunctions of the Hamiltonians associated with the commutative (integer ray) subalgebras of the Ding-Iohara-Miki algebra and those of the twisted Cherednik system. In the case of $t=q^{-m}$ with natural $m$, eigenfunctions of the first system of Hamiltonians are the twisted Baker-Akhiezer functions (BAFs) introduced by O. Chalykh, while eigenfunctions of the twisted Cherednik Hamiltonians are twisted non-symmetric Macdonald polynomials.
Actually, the twisted Cherednik ground state is symmetric and coincides with a peculiar symmetric BAF.
We lift this correspondence to excited states, and claim that
both Cherednik eigenfunctions and BAF's can be combined to produce symmetric functions, which coincide with each other and are eigenfunctions of the both DIM Hamiltonians and power sums of the twisted Cherednik Hamiltonians at once. This reflects the correspondence between the DIM algebra and the spherical DAHA explicitly.
\end{abstract}

\bigskip

\newcommand\smallpar[1]{
  \noindent $\bullet$ \textbf{#1}
}

\section{Introduction}

The Laplace operator in $\mathbb{R}^N$ is a sum of squares of commuting operators:
\be
\Delta = \left(\frac{\p}{\p \vec x}\right)^2
\ee
One can look for its eigenfunctions, and ask them to be  {\it symmetric} in all $x_i$.
It is not that immediate.
Instead, one can consider common eigenfunctions of gradient operators,
which is trivial, they are just exponentials
\be
\psi_{\vec p} = e^{i\vec p\vec x}
\ee
However, they are not symmetric in $x_i$.
In order to obtain a symmetric Laplace eigenfunction one can sum over many such functions,
say, over a sphere in the momentum space:
\be
{\cal Y}_{|\vec p|} = \int_{S^{d-1}} e^{i\vec p\vec x} d\vec p
\ee
The simplest realization of this idea is just
\be
Y_s=\sum_j e^{is x_j}
\ee
i.e. a sum of $N$ eigenfunctions with $N$ different eigenvalue sets
$(s,0,\ldots,0)$, $(0,s,\ldots,0)$ and $(0,0,\ldots,s)$
of $N$ commuting operators $-i\frac{\p}{\p x_i}$.
There is a variety of similar configurations, up to the peculiar monomial
$e^{is\sum_i x_i}$, which is the eigenfunction of all gradient operators at once, not only of the Laplace operator.

This simple example has a far-going generalization, relevant for the modern
theory of integrable systems.
The Laplace operator is substituted by a system of commuting Hamiltonians, for instance, the integrable Ruijsenaars-Schneider Hamiltonians $\hat H_k$ \cite{RS}, while the building blocks substituting gradient operators are the commuting Cherednik operators $\hat C_i$ \cite{Ch,NSCh} so that their power sums coincide with $\hat H_k$ on the space of symmetric functions:
\be
\hat H_k=\sum_i\hat C_i^k\Big|_{symm}
\ee
The Cherednik operators form a commuting system, and their common eigenfunctions form the system of non-symmetric polynomials $E_{w\lambda}(\vec x)$ parameterized by the \wcs \cite{Opd95,Mac96,Che95,HHL,MN,BF97}.

Hence, it does not come as a surprise that the symmetric eigenfunction
of $\hat H_k$ can be built from $E_{w\lambda}(\vec x)$ as a linear combination {\bf at fixed $\lambda$} with proper coefficients \cite{MN}.
These symmetric eigenfunctions are the standard symmetric Macdonald polynomials \cite{Mac}. In fact, they can be also obtained by symmetrization of the multivariable Baker-Akhiezer functions introduced by O. Chalykh in \cite{Cha}.

This general scheme can be described in the following picture:

\bigskip

\hspace{-1.4cm}
\fbox{\parbox{18.6cm}
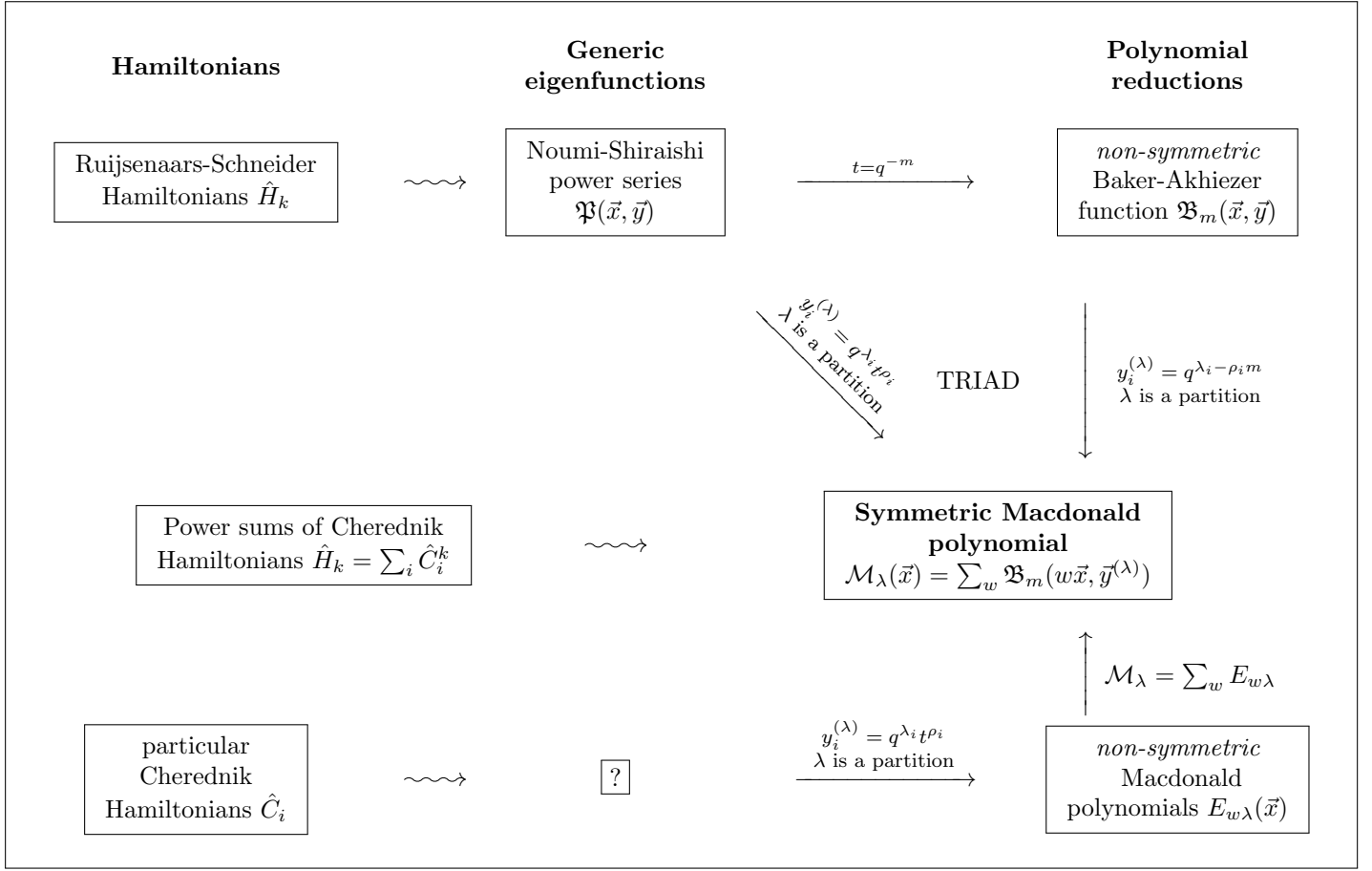
{
$$
\begin{array}{cccccr}
\hbox{\bf Hamiltonians}&&\begin{array}{c}\hbox{\bf Generic}\cr \hbox{\bf eigenfunctions}\end{array}&&
\begin{array}{c}\hbox{\bf Polynomial}\cr \hbox{\bf reductions}\end{array}&\cr
\cr
\boxed{\begin{array}{c}\hbox{Ruijsenaars-Schneider}\cr\hbox{Hamiltonians}\ \hat H_k
\end{array}}&\longsquiggly&\boxed{\begin{array}{c}
\hbox{Noumi-Shiraishi}\cr
\hbox{power series}\cr \mathfrak{P}(\vec x,\vec y)
\end{array}}&{\rotatebox[origin=c]{0}{\(\xrightarrow{\hspace*{.7cm}t=q^{-m}\hspace*{.7cm}}\)}}&
\boxed{\begin{array}{c}\hbox{\it non-symmetric}\cr\hbox{Baker-Akhiezer}\cr\hbox{function}\ \mathfrak{B}_m(\vec x,\vec y)
\end{array}}&
\cr
\\
&&&{\rotatebox[origin=c]{-45}{\(\xrightarrow{
{\footnotesize\begin{array}{c}
y_i^{(\lambda)}=q^{\lambda_i}t^{\rho_i}\cr\lambda\
\hbox{is a partition}
\end{array}}
}\)}}
\hbox{ TRIAD}
&{\rotatebox[origin=c]{-90}{\(\xrightarrow{\hspace*{2cm}}\)}}\ {{\footnotesize\begin{array}{c}
y_i^{(\lambda)}=q^{\lambda_i-\rho_im}\cr\lambda\
\hbox{is a partition}
\end{array}}}
\\
\cr
\hspace{1.5cm}
\boxed{\begin{array}{c}\hbox{Power sums of Cherednik}\cr\hbox{Hamiltonians}\ \hat H_k=\sum_i\hat C_i^k
\end{array}}
\hspace{-1.5cm}
&&\longsquiggly&& \hspace{-5.0cm}
\boxed{\begin{array}{c}\hbox{\bf Symmetric Macdonald}\cr\hbox{\bf polynomial}\cr {\cal M}_\lambda(\vec x )=\sum_w \mathfrak{B}_m(w\vec x,\vec y^{(\lambda)})\end{array}}
\cr
\\
&&&&{\rotatebox[origin=c]{-90}{\(\xleftarrow{\hspace*{1cm}}\)}}\ {\footnotesize{\cal M}_\lambda=\sum_w E_{w\lambda}
}
\\
\boxed{\begin{array}{c}\hbox{particular}\cr\hbox{Cherednik}\cr\hbox{Hamiltonians}\ \hat C_i
\end{array}}&\longsquiggly&\boxed{?}&{\rotatebox[origin=c]{0}{\(\xrightarrow{\footnotesize\begin{array}{c}
y_i^{(\lambda)}=q^{\lambda_i}t^{\rho_i}\cr\lambda\
\hbox{is a partition}
\end{array}}\)}}&
\boxed{\begin{array}{c}\hbox{\it non-symmetric}\cr\hbox{ Macdonald}\cr\hbox{polynomials}\ E_{w\lambda}(\vec x)
\end{array}}&
\cr
\end{array}
$$
}}

\bigskip

The three boxes at the upper right corner form what was called {\it triad} in \cite{triad}. It consists of the Noumi-Shiraishi power series, which is an eigenfunction of the Ruijsenaars-Schneider Hamiltonians at arbitrary $\vec y$ parameterizing the eigenvalues, and of two polynomial reductions: the Baker-Akhiezer function given at $t=q^{-m}$, $m\in\mathbb{Z}_{\ge 0}$ and the symmetric Macdonald polynomials emerging only at special values of $\vec y$ associated with partitions.

Question mark in the middle of the last line stands for a yet-unknown generalization of
the Noumi-Shiraishi functions, which will be non-symmetric and non-polynomial power series in $\vec x$
and also depend on the auxiliary $\vec y$ variable, with the $\vec x\leftrightarrow \vec y$ symmetry.

Once it gets found, we expect a new triad to be formed at the lower right corner of the figure.

\bigskip

Today of special interest is a further deformation of this picture: one starts with the Hamiltonians ${\cal H}_k^{(a)}$ that form the commutative subalgebra of the Ding-Iohara-Miki (DIM) algebra \cite{DI,Miki} associated with integer ray $(-1,a)$ \cite{MMP}.
The building blocks instead of gradient operators are $a$-twisted Cherednik operators $\hat {\mathfrak{C}}_i^{(a)}$ \cite{MMP}, and their power sums coincide with ${\cal H}_k^{(a)}$ on the space of symmetric functions \cite{MMP}:
\be
{\cal H}_k^{(a)}=\sum_i\left(\hat {\mathfrak{C}}_i^{(a)}\right)^k\Big|_{symm}
\ee
The $a$-twisted Cherednik operators form a commuting set, and their common eigenfunctions form a system of non-symmetric functions, which become polynomials at $t=q^{-m}$ ($m$ being natural): $E^{(a,m)}_{w\lambda}(\vec x)$
studied recently in \cite{MMP1,MMP2,MMP4}: in \cite{MMP1} have made a series of conjectures that have been proven in \cite{MMP4}.
It is therefore natural to ask a question, how (if any) the symmetric eigenfunctions
of ${\cal H}_k^{(a)}$ can be built from $E^{(a,m)}_{w\lambda}(\vec x)$. It turns out that one can generate symmetric functions as linear combinations of $E^{(a,m)}_{w\lambda}(\vec x)$ {\bf at fixed $\lambda$ with coefficients that do not depend on $a$}, i.e. with those used in the non-twisted case. Since, in the non-twisted case $a=1$,
these symmetric eigenfunctions are the ordinary Macdonald polynomials,
for an arbitrary $a$ we call them twisted symmetric Macdonald polynomials,
and denote then ${\cal M}^{(a)}_\lambda(\vec x)$.

Similarly to the non-twisted case, there is a twisted counterpart of the Baker-Akhiezer function \cite{ChE}, which is a non-symmetric eigenfunction of the Hamiltonians ${\cal H}_k^{(a)}$ \cite{ChF,MMPCha}. One can make the twisted symmetric Macdonald polynomials ${\cal M}^{(a)}_\lambda(\vec x)$ just by symmetrization of this function.

The whole picture in the twisted case looks as follows.

\bigskip

\hspace{-1.4cm}
\fbox{\parbox{18.6cm}
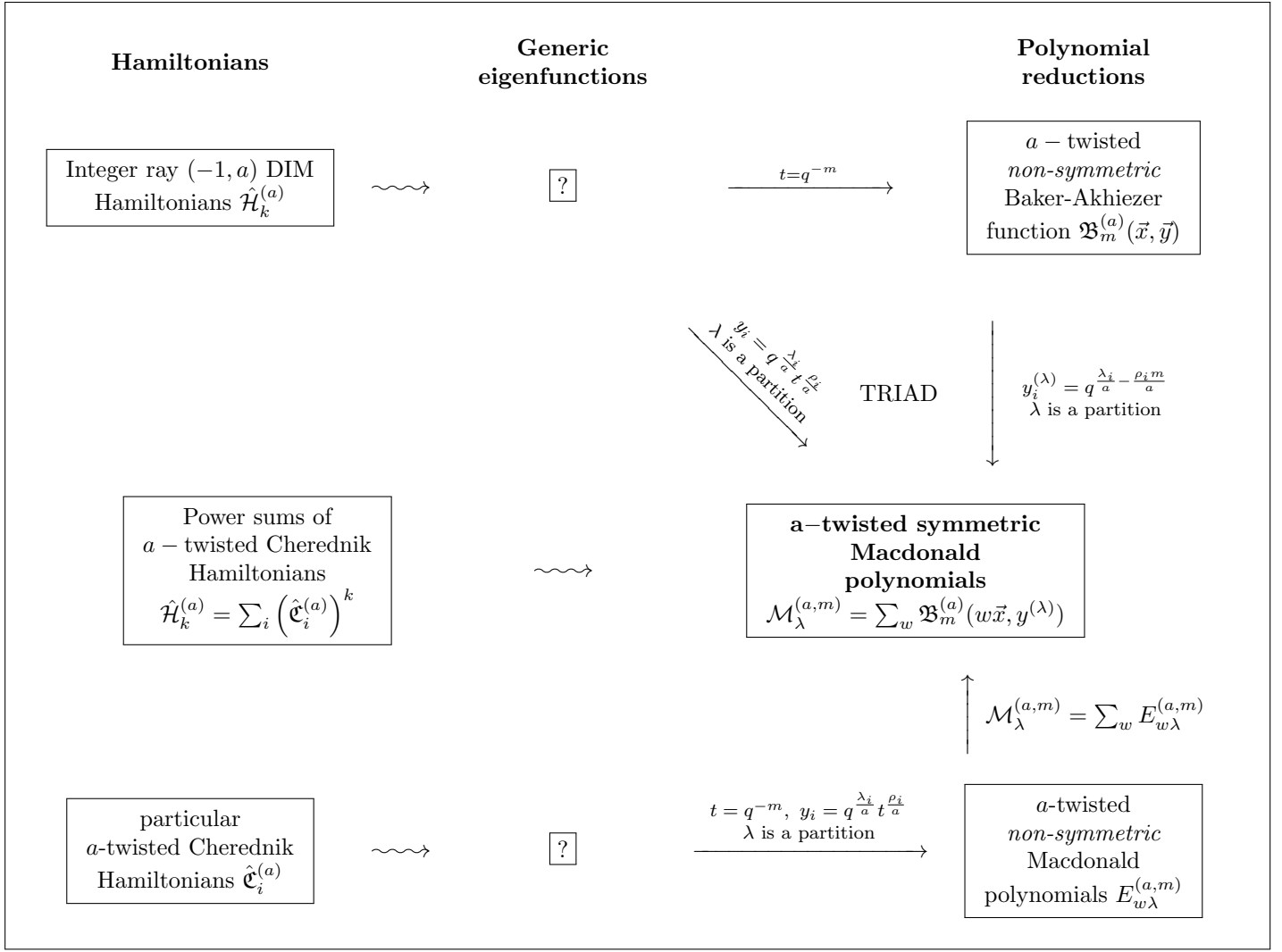
{
$$
\begin{array}{cccccr}
\hbox{\bf Hamiltonians}&&\begin{array}{c}\hbox{\bf Generic}\cr \hbox{\bf eigenfunctions}\end{array}&&
\begin{array}{c}\hbox{\bf Polynomial}\cr \hbox{\bf reductions}\end{array}&\cr
\cr
\boxed{\begin{array}{c}\hbox{Integer ray}\ (-1,a)\ \hbox{DIM}\cr\hbox{Hamiltonians}\ \hat {\cal H}_k^{(a)}
\end{array}}&\longsquiggly&\boxed{?}&{\rotatebox[origin=c]{0}{\(\xrightarrow{\hspace*{.7cm}t=q^{-m}\hspace*{.7cm}}\)}}&
\boxed{\begin{array}{c}a-\hbox{twisted}
\cr\hbox{\it non-symmetric}
\cr\hbox{Baker-Akhiezer}\cr\hbox{function}\ \mathfrak{B}_m^{(a)}(\vec x,\vec y)
\end{array}}&
\cr
\\
&&&{\rotatebox[origin=c]{-45}{\(\xrightarrow{{\footnotesize\begin{array}{c}\ y_i=q^{\lambda_i\over a}t^{\rho_i\over a}\cr\lambda\ \hbox{is a partition}\end{array}}}\)}}
\hbox{ TRIAD}
&{\rotatebox[origin=c]{-90}{\(\xrightarrow{\hspace*{2cm}}\)}}\ {{\footnotesize\begin{array}{c}
y_i^{(\lambda)}=q^{{\lambda_i\over a}-{\rho_im\over a}}\cr\lambda\
\hbox{is a partition}
\end{array}}}
\\
\cr
\hspace{1cm}\boxed{\begin{array}{c}\hbox{Power sums of}\cr a-\hbox{twisted Cherednik}\cr\hbox{Hamiltonians}\cr
\hat {\cal H}_k^{(a)}=\sum_i\Big(\hat {\mathfrak{C}}_i^{(a)}\Big)^k
\end{array}}
\hspace{-1cm}
&&\longsquiggly&&
\hspace{-5.0cm}
\boxed{\begin{array}{c}{\bf a-}\hbox{\bf twisted symmetric}\cr\hbox{\bf Macdonald}
\cr\hbox{\bf polynomials}\cr {\cal M}_\lambda^{(a,m)}=\sum_w \mathfrak{B}_m^{(a)}(w\vec x,y^{(\lambda)})\end{array}}
\cr
\\
&&&&{\rotatebox[origin=c]{-90}{\(\xleftarrow{\hspace*{1cm}}\)}}\ {\footnotesize{\cal M}_\lambda^{(a,m)}=\sum_w E_{w\lambda}^{(a,m)}
}
\\
\cr\boxed{\begin{array}{c}\hbox{particular}
\cr\hbox{$a$-twisted Cherednik}\cr\hbox{Hamiltonians}\ \hat {\mathfrak{C}}_i^{(a)}
\end{array}}&\longsquiggly&\boxed{?}&{\rotatebox[origin=c]{0}{\(\xrightarrow{
{\footnotesize\begin{array}{c}t=q^{-m},\ y_i=q^{\lambda_i\over a}t^{\rho_i\over a}\cr\lambda\ \hbox{is a partition}\end{array}}}
\)}}&
\boxed{\begin{array}{c}
\hbox{$a$-twisted}\cr\hbox{\it non-symmetric}\cr\hbox{Macdonald}\cr\hbox{polynomials}\ E_{w\lambda}^{(a,m)}
\end{array}}&
\cr
\end{array}
$$
}}

\bigskip

In the twisted case, there are two essential points: there is no known counterpart of the Noumi-Shiraishi function (see, however, \cite{MMP7} for particular results), and the polynomial non-symmetric eigenfunctions exist only upon the reduction $t=q^{-m}$, $m\in\mathbb{Z}_{>0}$, otherwise they are non-polynomial though sometimes could be associated with polynomials \cite[sec.5.4]{MMP1}. However, all other counterparts of the non-twisted case are present.

\bigskip

In this paper, we explain the details of this $a$-twisted picture. The paper is organized as follows: in section 2, we describe twisted DIM Hamiltonians and their eigenfunctions, which are the twisted Baker-Akhiezer functions. In section 3, we discuss the twisted Cherednik system and the corresponding eigenfunctions (twisted non-symmetric Macdonald polynomials), and, in section 4, we compare the two systems and construct from {\bf eigenfunctions of the twisted Cherednik system} the twisted symmetric Macdonald polynomials that are {\bf eigenfunctions of the DIM Hamiltonians}. This is our main claim, (\ref{main}). Section 5 contains some concluding remarks.

\paragraph{Notation.} Throughout the paper, we always put $t=q^{-m}$ with $m$ natural without special mentioning this. The reason is that the twisted Cherednik eigenfunctions are polynomial only in this case. These eigenfunctions are enumerated by \wcs similarly to the non-symmetric Macdonald polynomials. We mostly use the notation $w\lambda$ for them, which refers to a set of lines of the partition $\lambda$ (which has $\lambda_1\ge\lambda_2\ge\ldots\ge\lambda_N\ge 0$) under permutation $w$. We also use the notation $|\lambda|:=\sum_{i=1}^N\lambda_i$.

Throughout the paper, we freely use the term polynomial in the $a$-twisted case talking about polynomials of variables $x_i^{1\over a}$.

\section{DIM algebra}

\subsection{Commutative subalgebras of DIM algebra}

Commutative subalgebras of the DIM algebra are most explicit in the elliptic Hall algebra formulation of the DIM algebra. The elliptic Hall algebra is an associative algebra  multiplicatively generated by two central elements and elements $\mathfrak{e}_{\vec{\gamma}}$, with $\vec{\gamma} \in \mathbb{Z}^2\setminus \{(0,0)\}$, satisfying a set of commutation relations \cite{Feigin,BS,Zenk}. An important property of this algebra is that any vector $\vec\gamma$ gives rise to a commutative subalgebra:
\be\label{sb}
\left[ \mathfrak{e}_{\vec\gamma}, \mathfrak{e}_{k\vec \gamma}\right] = 0 \ \ \ \ \ \forall \vec\gamma \ {\rm and} \ k\in Z_+
\ee
The subalgebras associated with rays $\mathfrak{e}_{(\pm 1,a)}$ are called integer rays \cite{MMP}. In fact, all these subalgebras are related by the Miki automorphisms \cite{Miki1}, which represent action of the $SL(2,\mathbb{Z})$ group.

Various representations of the DIM algebra have been studied, here we are interested the $N$-body (or $N$-particle) representation of the algebra \cite{MMP}, which is just a tensor power of the vector representation \cite{Zenk2}. Commutative subalgebras in this representation give rise to integrable Hamiltonians of many-body systems, which generalize the trigonometric Ruijsenaars-Schneider systems.

We will discuss only integer rays $\mathfrak{e}_{(-1,a)}$, since the reflection symmetries: $\mathfrak{e}_{(k,m)}(x;q,t)\sim \mathfrak{e}_{(-k,m)}(x^{-1};q^{-1},t^{-1})$ and $\mathfrak{e}_{(k,m)}(q,t)=-\mathfrak{e}_{(k,-m)}(q^{-1},t^{-1})$ relate the rays in different quadrant of the $2d$ integer plane.

\subsection{Integer ray $N$-body Hamiltonians}

The $N$-body Hamiltonians associated with the integer ray $(-1,a)$ are symmetric with respect to permutations of $x_i$ and can be manifestly constructed following the procedure described in \cite{ChF},\cite[sec.6.2]{MMPCha},\cite[sec.7.6]{MMPformulas}. The Hamiltonians at $a=1$ are simply related with the trigonometric Ruijsenaars-Schneider Hamiltonians \cite{RS}, the first of them is
\be
\hat H^{(-1,1)}_1={\sqrt{q}\over q-1}\ \sum_{i=1}^N\prod_{j\ne i}{tx_i-x_j\over x_i-x_j}{1\over x_i}q^{\hat D_i}
\ee
with
\be
    \hat D_i:&=&x_i{\p\over\p x_i}
\ee
It is directly related with the Macdonald operator \cite{Mac,Macop}:
\be
\hat M=q^{-{1\over 2}\sum_{i=1}^n(\log_q x_i)^2}\cdot\hat H^{(-1,1)}_k\cdot q^{{1\over 2}\sum_{i=1}^n(\log_q x_i)^2}=
{\sqrt{q}\over q-1}\ \sum_{i=1}^N\prod_{j\ne i}{tx_i-x_j\over x_i-x_j}q^{\hat D_i}
\ee
The simplest example of the first Hamiltonian at $a=2$ is \cite[Eqs.(5.10)-(5.12)]{ChF},\cite[eq.(60)]{MMP}
\be\label{12N}
\hat H^{(-1,2)}_1&= & \ \sum_{i=1}^N \frac{1}{q^{1\over 2}x_i}
    \prod_{j \neq i} \frac{(t x_i - x_j)}{(x_i - x_j)}\frac{(qt x_i - x_j)}{(qx_i - x_j)}
    \ttop{i}\ttop{i}
    \nn \\
    & + &{q^{1\over 2}(t - q)(t - 1)\over q-1}
    \sum_{i \neq j} \prod_{k \neq i,j}
    \frac{(t x_i - x_k)(t x_j - x_k)}{(x_i - x_k)(x_j - x_k)}
    \frac{1}{(q x_i - x_j)} \ttop{i} \ttop{j}
\ee
Throughout the paper, we use the ``rotated'' Hamiltonians defined as
\be\label{rotH}
\hat {\cal H}^{(a)}_k=q^{-{1\over 2a}\sum_{i=1}^n(\log_q x_i)^2}\cdot\hat H^{(-1,a)}_k\cdot q^{{1\over 2a}\sum_{i=1}^n(\log_q x_i)^2}
\ee
for the reason that will be clear at the next subsection. Thus, $\hat {\cal H}^{(1)}_k$ is just $\hat M$.

\subsection{Eigenfunctions: twisted Baker-Akhiezer function}

The eigenfunction of the Hamiltonians $\hat H^{(-1,a)}_k$ is \cite{ChF,MMPCha} the twisted Baker-Akhiezer function $\mathfrak{B}_m^{(a)}(\vec x,\vec y)$ \cite{ChE}:
\be\label{efa}
\hat H^{(-1,a)}_k\cdot \left[q^{{1\over 2a}\sum_{i=1}^n(\log_q x_i)^2}\cdot\mathfrak{B}_m^{(a)}(\vec x,\vec y)\right]=
q^{-{amk\over 2}}\left(\sum_iy_i^k\right)\left[q^{{1\over 2a}\sum_{i=1}^n(\log_q x_i)^2}\cdot\mathfrak{B}_m^{(a)}(\vec x,\vec y)\right]
\ee
i.e.
\be\label{ef1}
\hat {\cal H}^{(-1,a)}_k\cdot \mathfrak{B}_m^{(a)}(\vec x,\vec y)=
q^{-{amk\over 2}}\left(\sum_iy_i^k\right)\cdot\mathfrak{B}_m^{(a)}(\vec x,\vec y)
\ee
The factors in formula (\ref{efa}) is the reason why we prefer to use the rotated Hamiltonians (\ref{rotH}).

The twisted Baker-Akhiezer function, which is a function of $2N$ complex parameters $x_i$ and $y_i$, $i=1,\ldots,N$, is defined as a sum
\be\label{BAtN}
\mathfrak{B}_m^{(a)}(\vec x,\vec y)=\left(\prod_{i=1}^Nx_i^{{\log_q y_i\over a}+m\rho_i}\right)\cdot \sum_{k_{ij}=0}^{ma}\prod_{i<j}\left({x_j\over x_i}\right)^{k_{ij}\over a}b^{(a)}_{m,\vec y,k}
\ee
with the property
\be
\mathfrak{B}_m^{(a)}(x_kq^j,\vec y)=\varepsilon^j\mathfrak{B}_m^{(a)}(x_lq^j,\vec y)\ \ \ \ \  \forall k,l\ \ \hbox{and}\ \ 1\le j\le m\ \ \ \ \ \hbox{at}\ \ \varepsilon x_k^{1\over a}= x_l^{1\over a}
\ee
for any $\varepsilon$ such that $\varepsilon^a=1$. Here $\vec\rho$ is the Weyl vector, i.e. $\rho_i={1\over 2}(N-2i+1)$. This $a$-twisted Baker-Akhiezer function is unique up to a normalization, and, upon a proper normalization, is symmetric with respect to the permutation of $\vec x$ and $\vec y$. However, we choose in this note another normalization:
\be\label{sym}
b^{(a)}_{m,\vec y,0}=1
\ee
This normalization matches the standard normalization of the Macdonald polynomials $P_\lambda$, while the symmetric one is designed for the bispectral (Macdonald-Ruijsenaars) duality \cite{Mac,Cha,NS} (see also a review in a wider context in \cite{MMdell}). Note that the generic Baker-Akhiezer function is a quasipolynomial because of the common monomial factor. However, we will be interested in this paper only in the polynomial Baker-Akhiezer functions.

\section{Twisted Cherednik system}

\subsection{Hamiltonians}

The commuting Hamiltonians of the twisted Cherednik \cite{Ch,NSCh} system are defined to be (see details in \cite{MMP1})
\be\label{sDAHA}
R_{ij}:&=&1+{(1-t^{-1})x_j\over x_i-x_j}(1-\sigma_{i,j})\\
R_{ij}^{-1}&=&1+{(1-t)x_j\over x_i-x_j}(1-\sigma_{i,j})\nn\\
{\cal C}_i^{(a)}&=&
t^{1 - i} \left(\prod_{j = i + 1}^n R_{i,j}\right)
    {1\over x_i^{a-1\over a}}q^{\hat D_i}
    \left(\prod_{j = 1}^{i - 1} R^{-1}_{j,i}\right)\nn\\
   \hat {\mathfrak{C}}_i^{(a)}:&=&{1\over x_i}\Big(x_i{\cal C}_i^{(a)}\Big)^a
\ee
 The products in ${\cal C}_i^{(a)}$ are arranged so that the smaller index stands to the left.

We put in these formulas $t=q^{-m}$, and are interested in the simultaneous eigenfunctions of these mutually commuting Hamiltonians:
\be\label{ef2}
\hat {\mathfrak{C}}_i^{(a)}\cdot E^{(a,m)}(\vec x)=\Lambda^{(a,m)}_i\cdot E^{(a,m)}(\vec x),\ \ \ \ \ \ \ i=1,\ldots,N
\ee

\subsection{Ground state eigenfunction}

The simplest of the eigenfunctions is the ground state, which is a symmetric function, as usual for the ground state eigenfunctions. Since symmetric eigenfunctions of the twisted Cherednik Hamiltonians $\hat {\mathfrak{C}}_i^{(a)}$ are simultaneously eigenfunctions of the DIM Hamiltonians ${\cal H}_k^{(a)}$ (see (\ref{HCh}) below), we use (\ref{ef1}) and note that this ground state eigenfunction in (\ref{ef2}) is $\mathfrak{B}_m^{(a)}(\vec x,\vec y)$ at some values of $\vec y$. One can see that these values are:
\be
y_i^{(\alpha)}=q^{(i-1)m+m{(a-2)(N-1)\over 2}+\alpha}
\ee
where $\alpha$ is an arbitrary constant. As soon as we need a polynomial eigenfunction, $\alpha=al$, where $l$ is a non-negative integer. At the ground state, $l=0$, i.e. $\alpha=0$. Thus, we finally obtain the ground state eigenfunction
\be
\Omega_m^{(a)}(\vec x)=\mathfrak{B}_m^{(a)}(\vec x,\vec y^{(0)})
\ee
This is a polynomial of variables $x_i^{1\over a}$ of degree $d_0=\frac{N(N-1)}{2}\cdot (a-1)\cdot m$.

At $a=1$, all $\Omega_m^{(1)}(\vec x)=1$. At $a=2$, $N=2$, it is given by a simple product:
\be\label{omega22}
\Omega_m^{(a)}(x_1,x_2)&=&\prod_{j=0}^{m-1}\Big(x_1^{1\over 2}+q^{j-\frac{m-1}{2}}x_2^{1\over 2} \Big)
\ee
At $N=2$ and arbitrary $a$, the expression is more involved \cite[Eq.(94)]{MMP1}, and generally the formulas for $\Omega_m^{(a)}(\vec x)$ become very tedious \cite{MMP2}.

\subsection{Twisted non-symmetric Macdonald polynomials}

The excitation eigenfunctions are all described basing on the ground state. They are enumerated by the weak compositions: one associates with any Young diagram $\lambda$ with lines $\lambda_1\ge\lambda_2\ge\ldots\ge\lambda_N\ge 0$, a set of eigenfunctions obtained by all possible permutations of the lines of $\lambda$. We will denote every permutation as $w\lambda$, with $w$ being an element of the symmetric group ${\cal S}_N$ (permutation). If one introduces the polynomial labeled by $\mu:=w\lambda$,
\be\label{Xi}
\Xi^{(a)}_{\mu}:=\left(\prod_{i=1}x_i^{\mu_i\over a}q^{\mu_i(\mu_i-1)\over 2a}\right)\ \Omega^{(a)}(q,t;\{q^{\mu_i}x_i\})=
\prod_{i=1}\left(x_i^{1\over a}q^{\hat D_i}\right)^{\mu_i}\ \Omega^{(a)}(q,t;\{x_i\})
\ee
then any polynomial eigenfunction in equation (\ref{ef2}) can be presented in the form of {\bf a linear combination of such $\Xi^{(a)}_{w\lambda}$ with coefficients that are rational functions independent on the twist $a$} \cite{MMP1,MMP4}:
\be\label{gE}
E^{(a,m)}_{w\lambda}(\vec x)=\sum_\nu F_{w\lambda,\nu}^{(m)}(\vec x)\cdot \Xi^{(a)}_{\nu}
\ee
where $F_{w\lambda,\nu}^{(m)}$ does not depend on $a$ at all (see \cite{MMP4} for an algorithmic procedure of their construction).

For instance, if $\lambda=[100]$, the three possible eigenfunctions are
\be\label{n3}
E_{[0,0,1]}^{(a,m)}(\vec x)&=&\Xi^{(a)}_{[001]}\\
E_{[0,1,0]}^{(a,m)}(\vec x)&=&{\Big\{{tx_2\over x_3}\Big\}\over \Big\{{x_2\over x_3}\Big\}}
\Xi^{(a)}_{[010]}+
{(1-t)\over 1-qt^2}{\Big\{{qt^2x_3\over x_2}\Big\}\over \Big\{{x_3\over x_2}\Big\}}\Xi^{(a)}_{[001]}\nn
\\
E_{[1,0,0]}^{(a,m)}(\vec x)&=&{\Big\{{tx_1\over x_2}\Big\}\Big\{{tx_1\over x_3}\Big\}\over \Big\{{x_1\over x_2}\Big\}\Big\{{x_1\over x_3}\Big\}}\Xi^{(a)}_{[100]}+
{(1-t)\over (1-qt)}{\Big\{{tx_2\over x_3}\Big\}\Big\{{qtx_2\over x_1}\Big\}\over \Big\{{x_2\over x_3}\Big\}\Big\{{x_2\over x_1}\Big\}}\Xi^{(a)}_{[010]}+{(1-t)\over (1-qt)}{\Big\{{tx_3\over x_2}\Big\}\Big\{{qtx_3\over x_1}\Big\}\over \Big\{{x_3\over x_1}\Big\}\Big\{{x_3\over x_2}\Big\}}\Xi^{(a)}_{[001]}\nn
\ee
where $t=q^{-m}$ and $\{x\}:=1-x$.

The eigenfunctions are polynomials of variables $x_i^{1\over a}$ of degree $d_\mu=d_0+|\lambda|$. The eigenvalues in (\ref{ef2}) at $\mu=\lambda$ are
\be\label{ev2}
\Lambda^{(a,m)}_i(\lambda)=q^{2(a-1)m+(i-1)m+{a-1\over 2}+\lambda_i}
\ee
and are obtained by permutations for other $\mu$, see (\ref{pev}) below.

A subtle point is the normalization of $E^{(a,m)}(\vec x)$, so far it was arbitrarily chosen. For our further consideration, we fix it in the following way. First of all, at $a=1$, we choose $\Omega^{(1)}(q,t;\{x_i\})=1$, the non-symmetric Macdonald polynomial becomes non-twisted, and it admits the triangle expansion
\be\label{Enorm}
E_{w\lambda}=x^{w\lambda}+\sum_{\mu,w}C_{\lambda\mu}x^{w'\mu}
\ee
where the sum runs over all diagrams $\mu<\lambda$ (in accordance with the lexicographic order), and, if $\mu=\lambda$, then the sum runs over all permutations $w'<w$ (in accordance with the Bruhat order). The normalization is fixed by the unit coefficient in front of $x^{w\lambda}$, and it determines the normalization of the function $F_{w\lambda,\nu}^{(m)}(\vec x)$ in (\ref{gE}).

\section{Correspondence between DIM and twisted Cherednik integrable systems}

\subsection{Correspondence between Hamiltonians}

In accordance with \cite[sec.6.2]{MMP}, the integer ray Hamiltonians ${\cal H}_k^{(a)}$ coincide with the power sums of the twisted Cherednik Hamiltonians $\hat {\mathfrak{C}}_i^{(a)}$ when acting on the space of symmetric functions:
\be\label{HCh}
{\cal H}_k^{(a)}\simeq\sum_i\left(\hat {\mathfrak{C}}_i^{(a)}\right)^k\Big|_{symm}
\ee
The sign $\simeq$ here implies that there are some simple normalization factors that are monomials in $q$. For the sake of brevity, we did not rescale the Hamiltonians to have the unit factors.

At the algebraic level, it expresses the relation \cite{DIMDAHA} between the DIM (or Elliptic Hall \cite{K,BS,S,Feigin}) algebra and spherical DAHA \cite{Ch}. Note that formula (\ref{HCh}) requires the very concrete normalization of the Hamiltonians $\hat {\mathfrak{C}}_i^{(a)}$ as in formula (\ref{sDAHA}). With this normalization of $\hat {\mathfrak{C}}_i^{(a)}$, there is the following basic property of
$E^{(a,m)}_\lambda(\vec x)$ \cite{MMP4}:
\be\label{pev}
\boxed{
\hat{\mathfrak{C}}_i^{(a)}\cdot E^{(a,m)}_{w\vec\lambda}(\vec x)=\Lambda^{(a,m)}_{(w\vec\lambda)_i}(\lambda)\cdot E^{(a,m)}_{w\vec\lambda}(\vec x)
}
\ee
This means that the eigenvalues for all $E^{(a,m)}_{w\vec\lambda}(\vec x)$ at fixed $\lambda$ are from the same set (\ref{ev2}), and choosing different $w$ just permutes the eigenvalues of different Hamiltonians $\hat{\mathfrak{C}}_i^{(a)}$. See details in \cite[sec.6.5]{MMP1} and in \cite{MMP4}.

For instance, in the case of $N=3$ for the excitation with $\lambda=[2]$, there are three \wcs, [200], [020] and [002]. When acting on the first of them, which corresponds to the Young diagram, one obtains from (\ref{ev2}) the eigenvalues
\be
\hat{\mathfrak{C}}_i^{(a)}\cdot E^{(a,m)}_{[200]}(\vec x)=\Lambda^{(a,m)}_i\cdot E^{(a,m)}_{[200]}(\vec x)\nn
\ee
\be
\Lambda^{(a,m)}_1&=&q^{2(a-1)m+{a-1\over 2}+2},\nn\\
\Lambda^{(a,m)}_2&=&q^{(2a-1)m+{a-1\over 2}},\nn\\
\Lambda^{(a,m)}_3&=&q^{2am+{a-1\over 2}}\nn
\ee
When acting by $\hat{\mathfrak{C}}_i^{(a)}$ on [020], the eigenvalues $\Lambda^{(a,m)}_1$ and $\Lambda^{(a,m)}_2$ are permuted, and when acting on [002], additionally permuted are the eigenvalues $\Lambda^{(a,m)}_2$ and $\Lambda^{(a,m)}_3$.

\subsection{Symmetric eigenfunctions: two realizations}

\subsubsection{Symmetric eigenfunction from the BAFs}

Though the BAF is not a symmetric polynomial, it can be transformed to a symmetric polynomial after summing up over all permutations:
\be
\Psi^{(a)}_m(\vec x,\vec y)=c_N\sum_{w\in W}\mathfrak{B}_m(w\vec x,\vec y)
\ee
where $W={\cal S}_N$ is the Weyl group, and $c_N$ is a numerical coefficient equal to ${1\over N!}$ when the BAF is a symmetric function of $x_i$ (i.e. one actually should not sum over the Weyl group), and equal to 1 otherwise.

Since
\be
\hat {\cal H}^{(-1,a)}_k\cdot \mathfrak{B}_m^{(a)}(\vec x,\vec y)=
q^{-{amk\over 2}}\left(\sum_iy_i^k\right)\cdot\mathfrak{B}_m^{(a)}(\vec x,\vec y)
\ee
and $H_k^{(a)}$ is symmetric with respect to permutations of $x_i$'s, one readily claims that
\be\label{HPsi}
\hat {\cal H}^{(-1,a)}_k\cdot\Psi^{(a)}_m(\vec x,\vec y)
=q^{-{amk\over 2}}\left(\sum_iy_i^k\right)\cdot \Psi^{(a)}_m(\vec x,\vec y)
\ee

\subsubsection{Symmetric eigenfunction from the twisted non-symmetric Macdonald polynomials}

Similarly, one can produce a symmetric twisted Macdonald polynomial by summing up the twisted non-symmetric Macdonald polynomials over the Weyl group acting on the \wc, i.e. over all permutations of the partition $\lambda$ \cite{MMP4} (see \cite{MN} for the non-twisted case):
\be\label{EM}
\boxed{
{\cal M}^{(a,m)}_{\lambda}(\vec x)=\sum_{{\mu=w\lambda}\atop{w\in W}} E^{(a,m)}_{\mu}(\vec x)\cdot\left(\prod_{(i,j):\ {{\mu_j>\mu_i}\atop{j>i}}}{1-q^{\mu_j-\mu_i}t^{\zeta(\mu)_i-\zeta(\mu)_j-1}\over
1-q^{\mu_j-\mu_i}t^{\zeta(\mu)_i-\zeta(\mu)_j}}\right)
}
\ee
i.e. the product in the summand runs over pairs of $(i,j)$ such that $\lambda_i<\lambda_j$ at $i<j$. The quantity $\zeta(\mu)_i$ here is defined for the\ \wc $\mu$ as
\be
\zeta(\mu)_i:=\#\{k<i|\mu_k\ge\mu_i\}+\#\{k>i|\mu_k>\mu_i\}
\ee

At $a=1$, $E^{(1,m)}_{\vec\lambda}(\vec x)$ is just the standard non-twisted non-symmetric Macdonald polynomial \cite{Opd95,Mac96,Che95}, and hence ${\cal M}^{(1,m)}_{\lambda}(\vec x)$ is the standard symmetric Macdonald polynomial \cite{Mac}.
Moreover, the coefficients in this sum do not depend on the twist $a$, and hence can be read off from the well-known case of standard $a=1$ non-symmetric Macdonald polynomials. Note that this formula is correct only when the twisted non-symmetric polynomials are normalized in the way described in formula (\ref{Enorm}).

A few words deserve adding to emphasize a non-triviality of this relation. Even at $a=1$,
the (non-twisted) non-symmetric Macdonald polynomials at level $\lambda$ depend on many parameters, much more than the number
of permutations of lines in the weak composition.
For instance, at $N=3$ the three \wcs, $[002]$, $[020]$ and $[200]$ depend on six monomials
$x_1^2,x_2^2,x_3^2,x_1x_2,x_1x_3,x_2x_3$.
Still, the three states can be combined to form a full symmetric expression:
\be
E_{[200]}+{(1-q^2)\over(1-q^2t)}E_{[020]}+{(1-q^2)\over(1-q^2t^2)}E_{[002]}
\nn
\ee
We adjusted coefficients of $x_i^2$ to get a symmetric expression, and the coefficients
of $x_ix_j$ are adjusted automatically, which was not {\it a priori} obvious.

This looks non-trivial even before twisting.
After twisting, the excitations are polynomials of much higher degree
$\frac{N(N-1)(a-1)m}{2}+|\lambda|$, only at $a=1$ they are reduced to degree $|\lambda|$.
Thus linear compositions are even more restricted.
Still there is a symmetric combination with the same coefficients!
For example, the combinations symmetric in all the three variables $x_1,x_2,x_3$ are:
\be
E_{[100]}^{(a)}+{(1-q)\over(1-qt)}E_{[010]}^{(a)}+{(1-q)\over(1-qt^2)}E_{[001]}^{(a)}, \nn \\
E_{[110]}^{(a)}+{(1-q)\over(1-qt)}E_{[101]}^{(a)}+{(1-q)\over(1-qt^2)}E_{[011]}^{(a)}, \nn \\
E_{[200]}^{(a)}+{(1-q^2)\over(1-q^2t)}E_{[020]}^{(a)}+{(1-q^2)\over(1-q^2t^2)}E_{[002]}^{(a)}
\ee
We remind that $E^{(a,m)}$ are polynomials in the variables $x_i^{1/a}$. As was claimed, the coefficients do not depend on $a$.
Note that, in the first two lines, the coefficients are the same, which is consistent with (\ref{EM}).

\subsubsection{Equivalence of two symmetric eigenfunctions}

It follows from formula (\ref{pev}) that {\bf any} linear combination of $E^{(a,m)}_{w\vec\lambda}(\vec x)$ with various $w$
is an eigenfunction of $\sum_i\left(\hat {\mathfrak{C}}_i^{(a)}\right)^k\Big|_{symm}$. In particular, this is the case for combination (\ref{EM}), which is a symmetric polynomial:
\be
\sum_i\left(\hat {\mathfrak{C}}_i^{(a)}\right)^k{\cal M}^{(a,m)}_{\lambda}(\vec x)
=\left[\sum_{i=1}^N\Big(\Lambda^{(a,m)}_i(\lambda)\Big)^k\right]\cdot{\cal M}^{(a,m)}_{\lambda}(\vec x)
\ee
where $\Lambda^{(a,m)}_i(\lambda)$ is given by (\ref{ev2}).

Since ${\cal M}^{(a,m)}_{\lambda}$ is a symmetric polynomial, (\ref{HCh}) implies that
\be
{\cal H}_k^{(a)}{\cal M}^{(a,m)}_{\lambda}(\vec x)
=q^{-{amk\over 2}}\left(\sum_i\Big(y_i^{(\lambda)}\Big)^k\right){\cal M}^{(a)}_{\lambda}(\vec x)
\ee
where
\be
y_i^{(\lambda)}:=q^{{\lambda_i\over a}-{m\over a}\rho_i}
\ee
On the other hand, $\Psi^{(a)}_m(\vec x,\vec y)$ is also a symmetric polynomial, and it is also satisfied equation (\ref{HPsi}).
Hence, we finally arrive at the equality
\be\label{main}
\boxed{
\Psi^{(a)}_m(\vec x,\vec y^{(\lambda)})\simeq{\cal M}^{(a)}_{\lambda}(\vec x)
}
\ee
where there can be a numerical coefficient. We checked in various examples that, with the normalization we have chosen, the coefficient is equal to 1!

In an expanded form, (\ref{main}) is
\be\label{ef}
\sum_{w\in W}\mathfrak{B}_m^{(a)}(w\vec x,\vec y^{(\lambda)})=\sum_{{\mu=w\lambda}\atop{w\in W}} E^{(a,m)}_{\mu}(\vec x)\cdot\left(\prod_{(i,j):\ {{\mu_j>\mu_i}\atop{j>i}}}{1-q^{\mu_j-\mu_i}t^{\zeta(\mu)_i-\zeta(\mu)_j-1}\over
1-q^{\mu_j-\mu_i}t^{\zeta(\mu)_i-\zeta(\mu)_j}}\right)
\ee

Moreover, we notice the equivariant property $\mathfrak{B}_m^{(a)}(w\vec x,w\vec y)=\mathfrak{B}_m^{(a)}(\vec x,\vec y)$ \cite{Cha,ChE} so that one can rewrite (\ref{ef}) in the form
\be
\boxed{
\sum_{w\in W}\mathfrak{B}_m^{(a)}(\vec x,w\vec y^{(\lambda)})=\sum_{{\mu=w\lambda}\atop{w\in W}} E^{(a,m)}_{\mu}(\vec x)\cdot\left(\prod_{(i,j):\ {{\mu_j>\mu_i}\atop{j>i}}}{1-q^{\mu_j-\mu_i}t^{\zeta(\mu)_i-\zeta(\mu)_j-1}\over
1-q^{\mu_j-\mu_i}t^{\zeta(\mu)_i-\zeta(\mu)_j}}\right)
}
\ee
However, though at the both sides of this equality the sums run over the Weyl group acting on $\vec y$, separate terms are different\footnote{This is evident since, for instance, separate terms at the l.h.s. celebrate the property $\mathfrak{B}_m^{(a)}(w\vec x,w\vec y)=\mathfrak{B}_m^{(a)}(\vec x,\vec y)$, while those at the r.h.s. do not.}: it is a kind of resummation formula.

\subsection{Examples}

In order to illustrate how these formulas work, we consider a couple of simple examples.

\paragraph{Example of $a=1$, $N=2$.} The simplest example is at $a=1$, $N=2$, and $\lambda=[2]$. We also require $m>1$, otherwise, there is a singularity. Then, \cite{Mac2}
\be
E_{[0,2]}^{(1,m)}&=&x_2^2+{1-t\over 1-qt}x_1x_2\nn\\
E_{[2,0]}^{(1,m)}&=&x_1^2+{q^2(1-t)\over 1-q^2t}x_2^2+{q(1+q)(1-t)\over 1-q^2t}x_1x_2
\ee
which, in accordance with (\ref{EM}), sums into the symmetric polynomial
\be
{\cal M}^{(1,m)}_{[2]}(x_1,x_2)=E_{[2,0]}^{(1,m)}+{(1-q^2)\over(1-q^2t)}E_{[0,2]}^{(1,m)}=x_1^2+x_2^2+{(1+q)(1-t)\over (1-qt)}x_1x_2
\ee
and \cite{MMPCha}
\be
\mathfrak{B}_m^{(1)}(x_1,x_2;q^{2-{m\over 2}},q^{m\over 2})=x_1^2+x_2^2+{(1+q)(1-q^{-m})\over (1-q^{1-m})}x_1x_2
\ee
In this case, it is symmetric\footnote{It becomes non-symmetric at higher $\lambda$: $|\lambda|\ge 2m+1$.} so that we obtain
\be
\Psi^{(1)}_m(x_1,x_2;q^{2-{m\over 2}},q^{m\over 2})=x_1^2+x_2^2+{(1+q)(1-q^{-m})\over (1-q^{1-m})}x_1x_2={\cal M}^{(1,m)}_{[2]}(x_1,x_2)
\ee

\paragraph{Examples of $a=2$, $N=2$.} Similarly, at $a=2$, $N=2$, and $\lambda=[2]$ \cite{MMP1}
\be
E_{[0,2]}^{(2,m)}&=&\Big(x_2+{\sqrt{qt}(1-t)\over(1-qt)}x_1^{1\over 2}x_2^{1\over 2}\Big)\Omega_m^{(a)}(x_1,x_2)\nn\\
E_{[2,0]}^{(2,m)}&=&\Big(x_1+
{q^2(1-t)\over (1-q^2t)}x_2+\sqrt{q\over t}{(1-t)(1-q^2t^2)\over (1-qt)(1-q^2t)}x_1^{1\over 2}x_2^{1\over 2}\Big)\Omega_m^{(a)}(x_1,x_2)\nn\\
\Omega_m^{(a)}(x_1,x_2)&=&\prod_{j=0}^{m-1}\Big(x_1^{1\over 2}+q^{j-\frac{m-1}{2}}x_2^{1\over 2} \Big)
\ee
which, in accordance with (\ref{EM}), sums into the symmetric polynomial
\be
{\cal M}^{(2,m)}_{[2]}(x_1,x_2)=E_{[2,0]}^{(2,m)}+{(1-q^2)\over(1-q^2t)}E_{[0,2]}^{(2,m)}=\Big(x_1+x_2+\sqrt{q\over t}{(1-t^2)\over (1-qt)}x_1^{1\over 2}x_2^{1\over 2}\Big)\Omega_m^{(a)}(x_1,x_2)
\ee
while \cite{MMPCha}
\be
\mathfrak{B}_m^{(2)}(x_1,x_2;q^{1-{m\over 4}},q^{m\over 4})=\Big(x_1+x_2+\sqrt{q\over t}{(1-t^2)\over (1-qt)}x_1^{1\over 2}x_2^{1\over 2}\Big)\prod_{j=0}^{m-1}\Big(x_1^{1\over 2}+q^{j-\frac{m-1}{2}}x_2^{1\over 2} \Big)
\ee
is again symmetric in this case, which gives rise to (\ref{main}):
\be
\Psi^{(2)}_m(x_1,x_2;q^{1-{m\over 4}},q^{m\over 4})={\cal M}^{(2,m)}_{[2]}(x_1,x_2)
\ee

\section{Conclusion}

It is well-known that the symmetric Macdonald polynomial can be constructed in two different ways: as a sum of non-symmetric Macdonald polynomials over permutations of lines of the Young diagram labelling this polynomial \cite{MN} and as a sum of multivariable Baker-Akhiezer functions over permutations of variables $x_i$ \cite{Cha}. At the algebraic level, this reflects the correspondence between the DIM algebra and the spherical DAHA. In this paper, we argued that the twisted non-symmetric polynomials can be also similarly constructed in two ways: as a sum of non-symmetric twisted Macdonald polynomials over permutations of lines of the Young diagram labelling this polynomial \cite{MMP1} and as a sum of multivariable Baker-Akhiezer functions introduced in \cite{ChE} over permutations of variables $x_i$. This is a corollary of the $SL(2,\mathbb{Z})$ symmetry, which allows one to produce twisted Hamiltonians from the standard Ruijsenaars and Cherednik ones (Miki automorphism \cite{Miki1,MMP}). We demonstrated that the whole construction is invariant under the twisting $SL(2,\mathbb{Z})$ transform.

However, there are still very essential points that should be understood within the framework described, these are marked by a few question marks in the figures in the Introduction. Besides, there is a question of more explicit description of the twisting $SL(2,\mathbb{Z})$ transform, which would allow one to generate all the results described here by directly applying this transform. There is also an issue of the more effective description of the twisted eigenfunctions, and of their extension to arbitrary values of $t$. We are looking forward at an advance in understanding these problems in the near future.

\section*{Acknowledgements}

The work was partially
funded within the state assignment of the Institute for Information Transmission Problems of RAS. Our work is also partly supported by Armenian SCS grants 24WS-1C031 (A.Mir.) and by the grant of the Foundation for the Advancement of Theoretical Physics and Mathematics ``BASIS''.

\end{document}